\documentclass[twocolumn]{article}
\usepackage[dvipdfmx]{graphicx}
\usepackage{pdfpages}
\usepackage{amsmath}
\usepackage{amssymb}
\usepackage{authblk}
\usepackage{bm}
\usepackage{braket}
\usepackage{lipsum}
\usepackage{float}

\title{Electrically-driven domain wall motion in a ferromagnetic Kagome lattice}
\author[1]{Sehoon Kim}
\author[2]{Daichi Kurebayashi}
\author[1,3]{Kentaro Nomura}
\affil[1]{Institute for Materials Research, Tohoku University, Sendai 980-8577, Japan}
\affil[2]{Center for Emergent Matter Science, RIKEN, Wako 351-0198, Japan}
\affil[3]{Center for Spintronics Research Network, Tohoku University, Sendai 980-8577, Japan}
\date{}

\begin{document}
\twocolumn[
\begin{@twocolumnfalse}
\maketitle
\begin{abstract}
We theoretically study domain wall motion induced by an electric field in the quantum anomalous Hall states on a two-dimensional Kagome lattice with ferromagnetic order and spin-orbit coupling. We show that an electric charge is accumulated near the domain wall which indicates that the electric field drives both the accumulated charge and the domain wall with small energy dissipation. Using the linear response theory we compute the non-equilibrium spin density which exerts a non-adiabatic spin transfer torque on textures of the local magnetization. This torque emerges even when the bulk is insulating and does not require the longitudinal electric current. Finally, we estimate the velocity of domain wall motion in this system, which is faster than that in conventional metals.
\end{abstract}
\end{@twocolumnfalse}
]
\indent  Spin torques are often used to control magnetic textures in spintronics as a new approach for memory devices, so-called spintronic devices. One of the proposals of spintronic devices uses the domain wall motion represented in the racetrack memory.\cite{Parkin190} Spin-polarized conduction electrons generate spin-transfer torques and induce the collective motion of local magnetic spins \cite{SLONCZEWSKI1996L1,PhysRevB.54.9353,TATARA2008213}. For device application, however, conventional ferromagnetic metals have Joule resistive heating problems caused by electric currents. Therefore, developing the less dissipative mechanism to control magnetic textures is demanded.\\
\indent  Topological materials with magnetic ordering are expected to be utilized in spintronic devices because of their dissipationless conduction on edges and domain walls. For instance, in a magnetic surface state on a three-dimensional topological insulator \cite{PhysRevB.82.161401,PhysRevB.81.241410,PhysRevLett.108.187201,PhysRevB.89.165307,PhysRevB.94.020411} and a Weyl semimetal \cite{Kurebayashi_Weyl_SciRep}, controlling magnetic textures by utilizing topological edge states has been theoretically proposed. It is shown that those proposals have much higher efficiency than that of the conventional metallic ferromagnets.\\
\begin{figure}[h] 
\centering
\includegraphics[width=7.5cm]{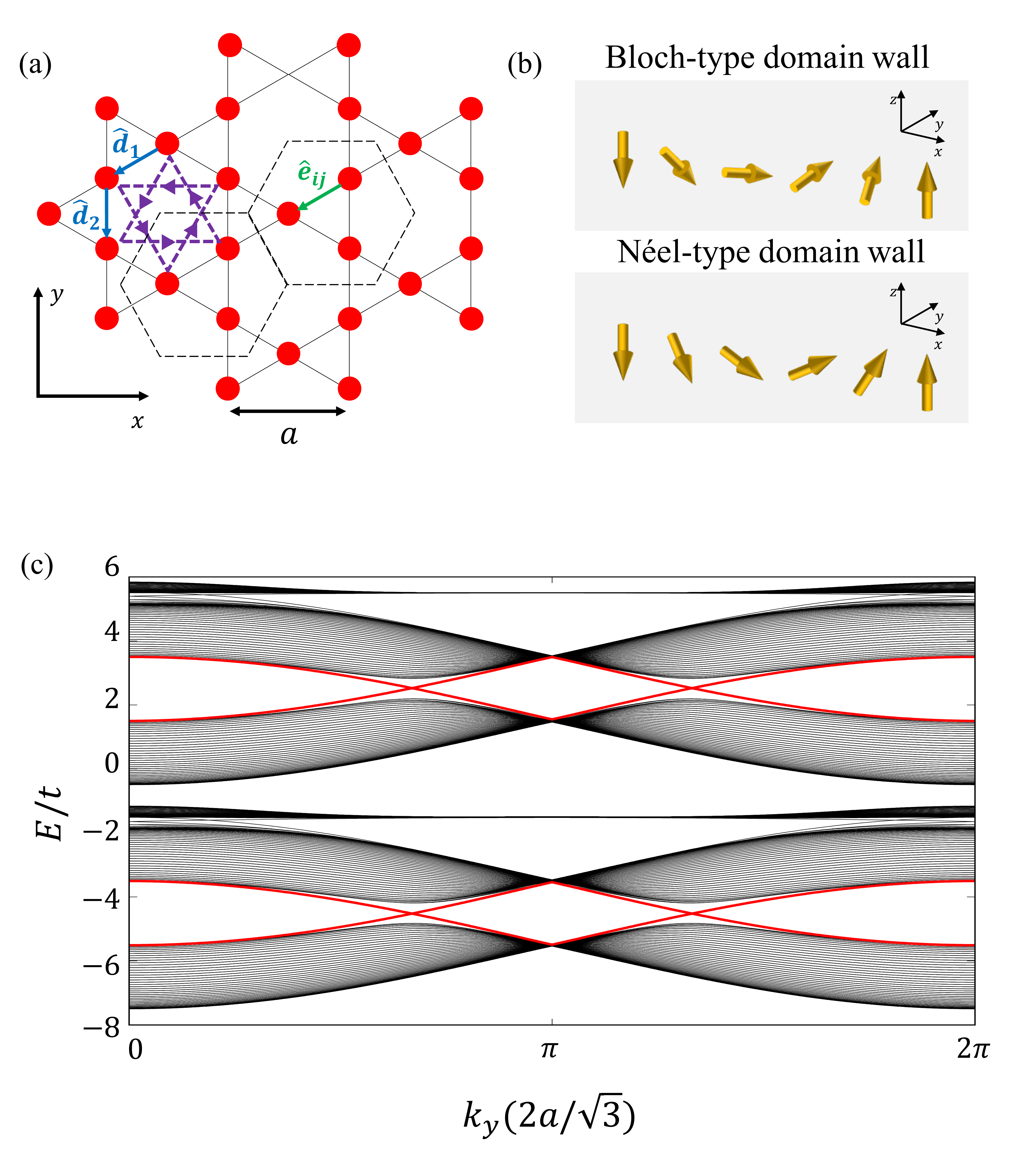}
\caption[]{(Color online) A schematic diagram of (a) a two-dimensional Kagome lattice and (b) its magnetic texture with a domain wall: Bloch-type and N\'eel-type domain wall. Dotted line and blue arrow line in (a) denote the unit cell and a nearest-neighbor vector, respectively. Second-nearest neighbor hopping amplitude becomes \begin{math}+i\lambda_{\mathrm{so}}\end{math} along the purple arrow and \begin{math}-i\lambda_{\mathrm{so}}\end{math} against the arrow. \begin{math}\hat{\bm{d}}_{1,2}\end{math}. \begin{math}a\end{math} is a lattice constant. (c) A band structure with the domain wall where red lines and \begin{math}E_{\mathrm{D}}\end{math} denote edge states around domain walls.}\label{Kagome}
\end{figure}
\indent Quantum anomalous Hall state, \cite{PhysRevLett.61.2015,PhysRevLett.90.206601,PhysRevLett.101.186807,Yu61,PhysRevB.82.161414} which has a quantized Hall conductivity without an external magnetic field, is achieved by ferromagnetic order and strong spin-orbit coupling. It is different from the quantum spin Hall insulator because of broken time-reversal symmetry. The value of the Hall conductivity is given by the Chern number leading to the chiral edge conduction when the bulk is insulating. Experimentally, the quantum anomalous Hall effect was observed at two-dimensional surface of three-dimensional topological insulators. \cite{Chang167} It is known that chiral edge modes are realized at the boundary between two quantum anomalous Hall states with different Chern numbers, such as magnetic domain walls.\cite{Wakatsuki_2015,Yasuda1311} Since a finite electric charge is accumulated near the domain walls, the motion of domain walls is expected be driven by an electric field. The experimental condition to observe those phenomena, however, requires low temperature under 1 K in the case of magnetic topological insulators.\\
\indent Meanwhile, a ferromagnetic Kagome lattice \cite{PhysRevLett.115.186802,PhysRevB.62.R6065,PhysRevB.80.113102,Zhang_JPhys_2011,PhysRevLett.112.017205,Fujiwara_2019} has also been considered as a candidate of the quantum anomalous Hall state. It has similar topological features with a honeycomb lattice with the spin-orbit coupling and a ferromagnetic order, which is known as the Haldane model.\cite{PhysRevLett.61.2015,PhysRevLett.95.146802} Contrary to the honeycomb structure, the Kagome lattice has one flat and two dispersive bands which intersect at two different Dirac points. Furthermore intrinsic spin-orbit coupling creates a band gap and the nontrivial Chern number. If the chemical potential is located in the Dirac gap, the quantum anomalous Hall effect occurs.\\
\indent Recently, several materials with higher Curie temperature are reported as the quantum anomalous Hall state experimentally. Ye \textit{et al.} \cite{Ye_Nature_2018} report the Dirac gap induced by the spin-orbit coupling in a \begin{math}\mbox{Fe}_{3}\mbox{Sn}_{2}\end{math} Kagome-lattice metal at room temperature. Liu \textit{et al.}  \cite{Liu_2018} show the giant anomalous Hall effect in  \begin{math}\mbox{Co}_{3}\mbox{Sn}_{2}\mbox{S}_{2}\end{math} which is a magnetic Weyl semimetal with the easy axis of the magnetization lying along out-of-plane even with the relatively high Curie temperature \begin{math}T_{C}\simeq 177 \mbox{ K}\end{math}. Because the magnetic Weyl semimetal is regarded as stacked heterostructures of the quantum anomalous Hall state and a spacer, \cite{PhysRevLett.107.127205} the quantum anomalous Hall state is expected to appear in a two-dimensional limit of \begin{math}\mbox{Co}_{3}\mbox{Sn}_{2}\mbox{S}_{2}\end{math} \cite{2017arXiv171208115M}.\\
\indent In this letter, we investigate an electrically-induced spin torque in the ferromagnetic Kagome lattice. We calculate a spin density \begin{math}\delta\langle\hat{\mbox{\boldmath $\sigma$}}\rangle\end{math} as a response \cite{PhysRevB.64.014416,PhysRevB.72.245330,garate_transport_2009} to the external electric field. Consequently, the non-adiabatic spin torque is obtained around the domain wall. Furthermore, we study the domain wall motion.\\
\indent The model we study is the two-dimensional Kagome lattice with a ferromagnetic order [Fig. \ref{Kagome}(a)]. The intrinsic spin-orbit coupling and the exchange interaction between local spins and conduction electron spins are needed to make a ferromagnetic Kagome lattice the quantum anomalous Hall state. In our work, we introduce the spin-dependent second-nearest neighbor hopping acting as the intrinsic spin-orbit coupling. The Hamiltonian is \cite{PhysRevB.80.113102,PhysRevLett.95.146802}
\begin{align} \label{eq:H}
H
&=
-t\sum_{\langle ij \rangle\alpha}\hat{c}^{\dagger}_{i\alpha}\hat{c}_{j\alpha}+i\lambda_{\mathrm{so}}\sum_{\langle\langle ij \rangle\rangle\alpha\beta}\nu_{ij}\hat{c}^{\dagger}_{i\alpha}(\hat{\sigma}_{z})_{\alpha\beta}\hat{c}_{j\beta}\nonumber \\
&+i\lambda_{\mathrm{R}}\sum_{\langle ij \rangle\alpha\beta}(\hat{\mbox{\boldmath $\sigma$}}_{\alpha\beta}\times\hat{\bm{e}}_{ij})_{z}\hat{c}^{\dagger}_{i\alpha}\hat{c}_{j\beta} \nonumber \\
&-J_{\mathrm{exc}}S\sum_{i\alpha\beta}(\hat{\bm{M}}\cdot\hat{\mbox{\boldmath $\sigma$}})_{\alpha\beta}\hat{c}^{\dagger}_{i\alpha}\hat{c}_{j\beta},
\end{align}
where \begin{math}\hat{c}_{i\alpha}^{\dagger}\end{math} denotes an electron creation operator with the spin \begin{math}\alpha=\uparrow,\downarrow\end{math} at \begin{math}i\end{math} site. The first term is a nearest neighbor hopping term with a hopping amplitude \begin{math}t\end{math}. The second term is a intrinsic spin-orbit coupling with the mirror symmetry and \begin{math}\nu_{ij}=(2/\sqrt{3})(\hat{\bm{d}}_{1}\times\hat{\bm{d}}_{2})=\pm 1\end{math} is the spin-dependent second-nearest neighbor hopping where \begin{math}\hat{\bm{d}}_{1,2}\end{math} are nearest-neighbor vectors traversed between second neighbor \begin{math}i\end{math} and \begin{math}j\end{math}. \begin{math}\lambda_{\mathrm{so}}\end{math} is an intrinsic spin-orbit coupling constant. The third term is a nearest-neighbor Rashba term with the amplitude \begin{math}\lambda_{\mathrm{R}}\end{math} and a nearest neighbor vector \begin{math}\hat{\bm{e}}_{ij}\end{math}. It is formed due to the \begin{math}+z\leftrightarrow -z\end{math} mirror symmetry breaking such as the interaction with a substrate. \begin{math}J_{\mathrm{exc}}\end{math} is an exchange coupling constant with local spins which have an amplitude \begin{math}S\end{math} and a directional vector \begin{math}\hat{\bm{M}}\end{math}. When the ferromagnetic order is the Stoner type, as in the low spin states, the local magnetization field \begin{math}\hat{\bm{M}}\end{math} should be regarded as the order parameter field in the mean-field approximation.\cite{2019arXiv190408148O} When \begin{math}J_{\mathrm{exc}}\end{math} is large enough the system becomes the fully spin-polarized quantum anomalous Hall insulator. In this work we consider a nanoribbon geometry on the Kagome lattice, with a periodic boundary condition imposed in \begin{math}y\end{math}-direction.\\
\indent Supposing the system has inhomogeneous magnetic moments with domain walls [Fig. \ref{Kagome}(b)], we set a spatial distribution along \begin{math}x\end{math} direction of local spins \begin{math}\hat{\bm{M}}\rightarrow\hat{\bm{M}}(x)\end{math}. In the uniaxial magnetic anisotropy, the expression of the domain wall is given by \begin{math}\hat{\bm{M}}=(\sin\theta(x)\cos\phi,\sin\theta(x)\sin\phi,\cos\theta(x))\end{math} where \begin{math}\theta(x)=2\arctan[\exp\{-\pi x/\xi\}]\end{math}, \begin{math}\phi=\pi/2,3\pi/2\end{math} represent Bloch-type domain walls and \begin{math}\phi=0,\pi\end{math} represent the N\'eel-type domain walls. \begin{math}\xi\end{math} is a domain wall width. We define dimensionless \begin{math}x_{a}\end{math} and \begin{math}\xi_{a}\end{math} as the position and the domain wall width divided by the lattice constant \begin{math}a\end{math}, respectively. As seen in the band structure with the domain wall [Fig. \ref{Kagome}(c)], the edge states around the domain wall appear inside the bulk gap.\\
\begin{figure}[h] 
\centering
\includegraphics[width=8cm]{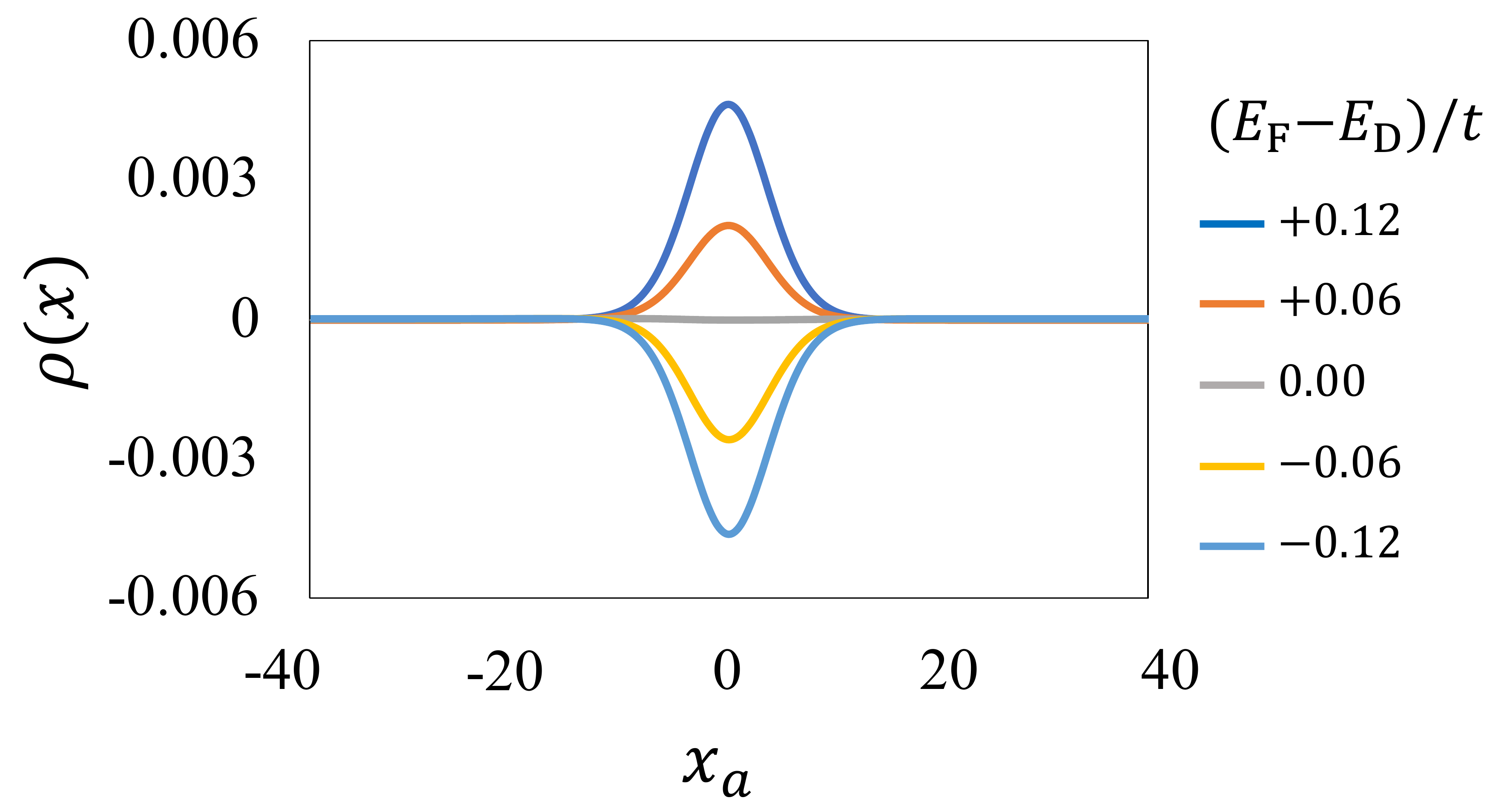}
\caption[]{(Color online) Local electron density \begin{math}\rho(x)\end{math} in the presence of a domain wall for various fermi energy \begin{math}E_{\mathrm{F}}\end{math} in a band gap where \begin{math}E_{\mathrm{D}}\end{math} denotes the 1/6-filling point. We choose the Hamiltonian parameters as \begin{math}\lambda_{\mathrm{so}} / t = 0.1\end{math}, \begin{math}\lambda_{\mathrm{R}}/t=0\end{math}, \begin{math}J_{\mathrm{exc}}S / t = 3.5\end{math}, and \begin{math}\xi_{a} = 30\end{math}.} \label{ED}
\end{figure}
\indent We show an electron density distribution (Fig. \ref{ED}) around the domain wall via a numerical calculation
\begin{equation}
\rho(x)=\sum_{k_{y},n\in \mathrm{occupied}}(|\psi_{k_{y}}^{\mathrm{DW}}(x)|^{2}-|\psi_{k_{y}}^{\mathrm{uniform}}(x)|^{2}),
\end{equation}
where \begin{math}\psi_{k_{y}}^{\mathrm{DW}}(x)\end{math} and \begin{math}\psi_{k_{y}}^{\mathrm{uniform}}(x)\end{math} denote eigenstates under the domain wall texture and the uniform local spins, respectively. Due to edge conductions of each magnetic domain, charges are localized around the domain wall. We expect that these localized charges are influenced by the external electric field, leading to the domain wall motion.\\
\begin{figure*}[h] 
\centering
\includegraphics[width=14cm]{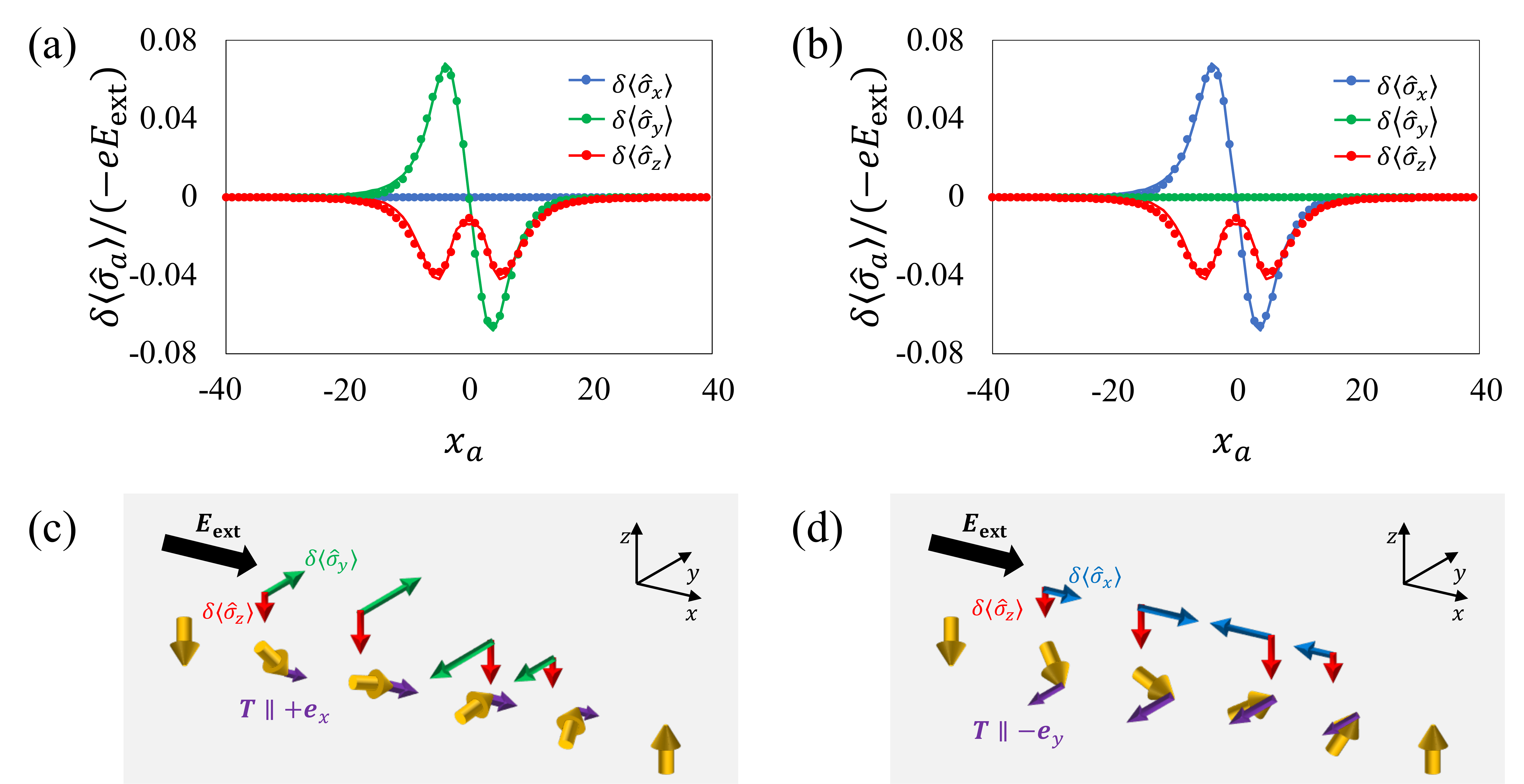}
\caption[]{(Color online) Spin accumulation \begin{math}\delta\langle\hat{\mbox{\boldmath $\sigma$}}(x)\rangle\end{math} without the Rashba spin-orbit coupling (\begin{math}\lambda_{\mathrm{R}}/\lambda_{\mathrm{so}}=0\end{math}) under (a) Bloch-type domain wall and (b) N\'eel-type domain wall. Dots and solid lines in each figure denote numerical results from Eq. (\ref{eq:LRT}) and fitting results from Eq. (\ref{eq:spina}), respectively. For fitting process, we use the summation of Eq. (\ref{eq:spina}) until \begin{math}n=2\end{math}. We choose the Hamiltonian parameters as \begin{math}\lambda_{\mathrm{so}} / t = 0.1\end{math}, \begin{math}\lambda_{\mathrm{R}}/t=0\end{math}, \begin{math}J_{\mathrm{exc}}S / t = 3.5\end{math}, and \begin{math}\xi_{a} = 30\end{math}. Fermi energy is \begin{math}E_{F}/t=-4.6\end{math}. Schematic diagrams of the electrically-induced spin torque \begin{math}\bm{T}\end{math} around (c) the Bloch-type and (d) N\'eel-type domain wall. Yellow arrows denote magnetic textures.} \label{spin}
\end{figure*}
\indent To describe magnetization dynamics, Landau-Lifshitz-Gilbert (LLG) equation is used:
\begin{equation} \label{eq:LLG}
\frac{d\hat{\bm{M}}(x)}{dt}=-\gamma\hat{\bm{M}}(x)\times\bm{H}_{\mathrm{ext}}(x)+\alpha\hat{\bm{M}}(x)\times\frac{d\hat{\bm{M}}(x)}{dt}+\bm{T},
\end{equation}
where \begin{math}\gamma\end{math} is a gyromagnetic ratio, \begin{math}\bm{H}_{\mathrm{ext}}\end{math} is an external magnetic field, and \begin{math}\alpha\end{math} is a dimensionless damping constant. An additional term \begin{math}\bm{T}\end{math} corresponds to a spin torque, which is generated by the spin polarization \begin{math}\delta\langle\hat{\mbox{\boldmath $\sigma$}}(x)\rangle\end{math} of itinerant electrons. The torque \begin{math}\bm{T}\end{math} acting on \begin{math}\hat{\bm{M}}\end{math} is expressed by an outer product of \begin{math}\hat{\bm{M}}\end{math} and \begin{math}\delta\langle\hat{\mbox{\boldmath $\sigma$}}(x)\rangle\end{math} as follows \cite{PhysRevLett.93.127204,Kohno.JPSJ.2006}
\begin{equation} \label{eq:STT}
\bm{T}=\frac{J_{\mathrm{exc}}S}{\hbar\rho_{S}}\hat{\bm{M}}\times\delta\langle\hat{\mbox{\boldmath $\sigma$}}(x)\rangle,
\end{equation}
where \begin{math}\rho_{S}\end{math} is the number of local magnetic moments per unit volume. The spin polarization \begin{math}\delta\langle\hat{\mbox{\boldmath $\sigma$}}(x)\rangle\end{math} is obtained by the linear response theory as \cite{PhysRevB.64.014416,PhysRevB.72.245330,garate_transport_2009}
\begin{align}
\frac{\delta\langle\hat{\mbox{\boldmath $\sigma$}}(x)\rangle}{-eE_{\mathrm{ext}}}
&=
\frac{i\hbar}{\Omega}\sum_{k_{y},a,b}\frac{f(\epsilon_{k_{y},a})-f(\epsilon_{k_{y},b})}{\epsilon_{k_{y},b}-\epsilon_{k_{y},a}-i\eta}\nonumber \\
&\times
\frac{\langle k_{y},a |\hat{\mbox{\boldmath $\sigma$}}(x)|k_{y},b\rangle\langle k_{y},b|\hat{v}_{x}|k_{y},a\rangle}{\epsilon_{k_{y},b}-\epsilon_{k_{y},a}}, \label{eq:LRT}
\end{align}
where \begin{math}a\end{math}, \begin{math}b\end{math} are band indices, \begin{math}\epsilon_{k_{y},n}\end{math} is an energy of a state \begin{math}|k_{y},n\rangle\end{math} (\begin{math}n=a,b\end{math}) and \begin{math}\hat{v}_{x}\end{math} is a \begin{math}x\end{math} component of a velocity operator.\\
\indent In conventional metals, a spin-transfer torque \begin{math}\bm{T}_{\mathrm{STT}}\end{math} with the lowest order in \begin{math}\hat{\bm{M}}\end{math} is given by \cite{Kohno.JPSJ.2006}
\begin{equation} \label{eq:totaltorque}
\bm{T}_{\mathrm{STT}}=(\bm{a}\cdot\mbox{\boldmath $\nabla$})\hat{\bm{M}}+\hat{\bm{M}}\times(\bm{b}\cdot\mbox{\boldmath $\nabla$})\hat{\bm{M}},
\end{equation}
where \begin{math}\bm{a}\end{math}-term is an adiabatic spin-transfer torque and \begin{math}\bm{b}\end{math}-term is a non-adiabatic spin-transfer torque. This expression can be expanded to higher orders in the spatial gradient of \begin{math}\hat{\bm{M}}\end{math} to compare with experimental results. \cite{Hals_prb_spin} The adiabatic spin-transfer torque is based on the exchange interaction between electron spins and local spins. The non-adiabatic contribution could be stronger than the adiabatic contribution in the presence of a spin-orbit coupling system causing a spin relaxation.\cite{TATARA2008213} Therefore we expect the electrically-induced non-adiabatic torque is predominant over the adiabatic torque in the quantum anomalous Hall state rather than in conventional metals.\\
\indent Here, we calculate the electrically-induced spin polarization \begin{math}\delta\langle\hat{\mbox{\boldmath $\sigma$}}(x)\rangle\end{math} on the domain wall without the Rashba spin-orbit coupling (\begin{math}\lambda_{\mathrm{R}}/\lambda_{\mathrm{so}}=0\end{math}). Note that the Fermi energy \begin{math}E_{F}\end{math} is inside the bulk gap. As we expected, the spin polarization is localized around domain wall. Figure \ref{spin} shows the spin density in cases of both (i) the Bloch-type domain wall [Fig. 3(a)] and (ii) the N\'eel-type domain wall [Fig. 3(b)]. The perpendicular component \begin{math}\delta\langle\hat{\sigma}_{\perp}(x)\rangle\end{math} depends on the domain wall type, while  \begin{math}z\end{math} component \begin{math}\delta\langle\hat{\sigma}_{z}(x)\rangle\end{math} does not.  From results, we obtain phenomenological expressions of the spin accumulation \begin{math}\delta\langle\hat{\mbox{\boldmath $\sigma$}}(x)\rangle\end{math} with the lowest order in \begin{math}\hat{\bm{M}}\end{math} as follows
\begin{equation}
\delta\langle\hat{\sigma}_{a}(x)\rangle
=
\sum_{n=0}^{\infty}\chi_{a}^{(2n+1)}\frac{d^{2n+1}}{dx^{2n+1}}\hat{M}_{a}(x)E_{\mathrm{ext}} \quad (a=\perp,z), \label{eq:spina}
\end{equation}
where \begin{math}\chi_{a}\end{math} is a coefficient which characterizes a magnitude of a response from the electric field. We illustrate the direction of the spin density and the induced spin torque in Figs. \ref{spin}(c) and \ref{spin}(d). They correspond to the spin polarization of localized charges around the domain walls. Comparing Eq. (\ref{eq:totaltorque}) with Eq. (\ref{eq:spina}), this electrically-induced spin accumulation causes the non-adiabatic spin-transfer torque. This non-adiabaticity induces \begin{math}\dot{\phi}\end{math} through the competition between \begin{math}\delta\langle\hat{\sigma}_{\perp}(x)\rangle\end{math} and \begin{math}\delta\langle\hat{\sigma}_{z}(x)\rangle\end{math}. The appearance of the non-adiabatic torque makes the intrinsic pinning effect\cite{PhysRevLett.92.086601} vanish, causing the disappearance of the intrinsic threshold current for the domain wall motion.\\
\begin{figure}[h] 
\centering
\includegraphics[width=7cm]{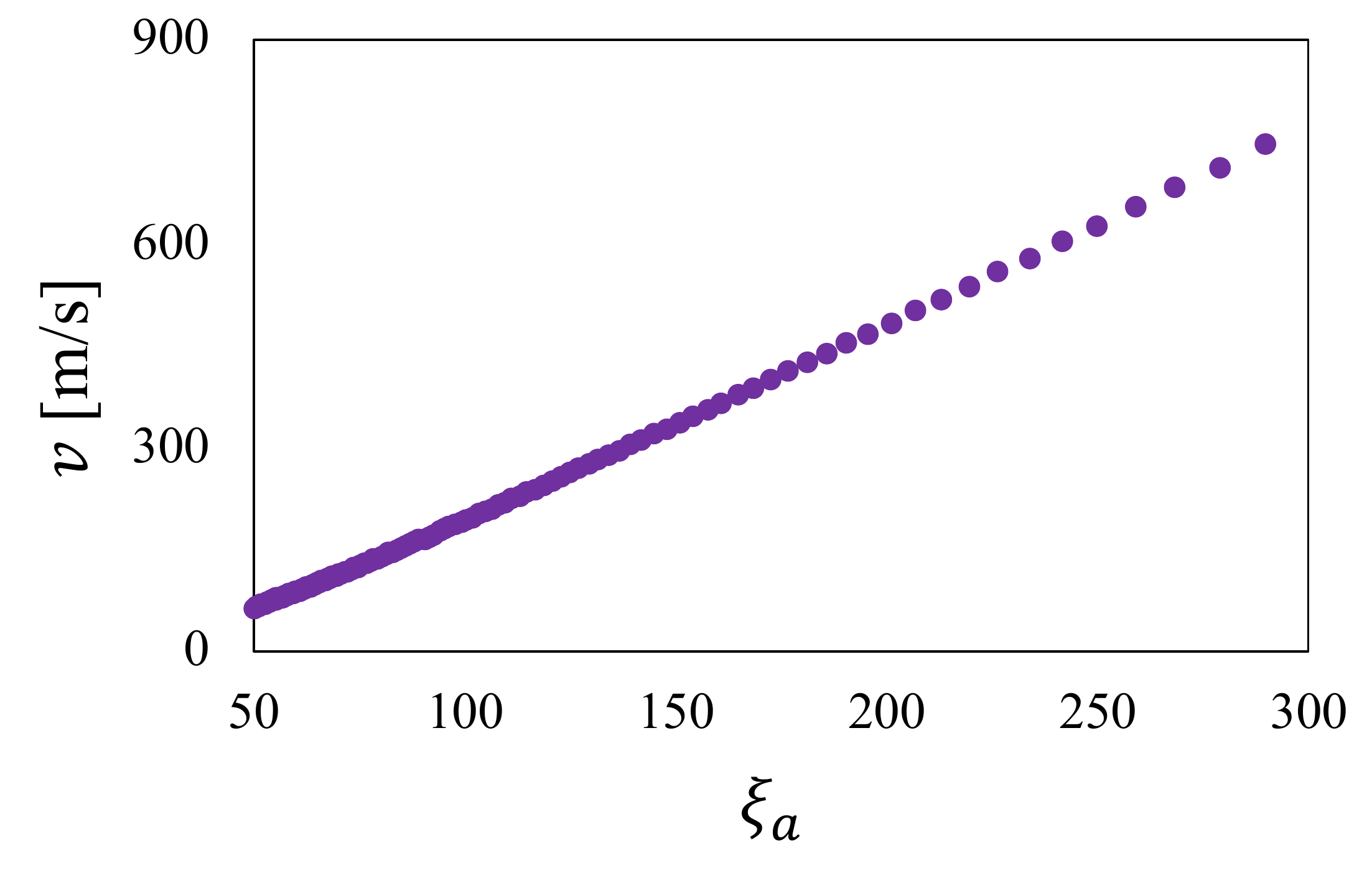}
\caption[]{(Color online) Domain wall width \begin{math}\xi_{a}\end{math}-dependence of the domain wall velocity \begin{math}v\end{math}. We choose the values of parameters as \begin{math}J_{\mathrm{exc}}S=0.2\end{math} eV and \begin{math}E=10^{5}\end{math} V/m.} \label{velo}
\end{figure}
\begin{figure*}[h] 
\centering
\includegraphics[width=14cm]{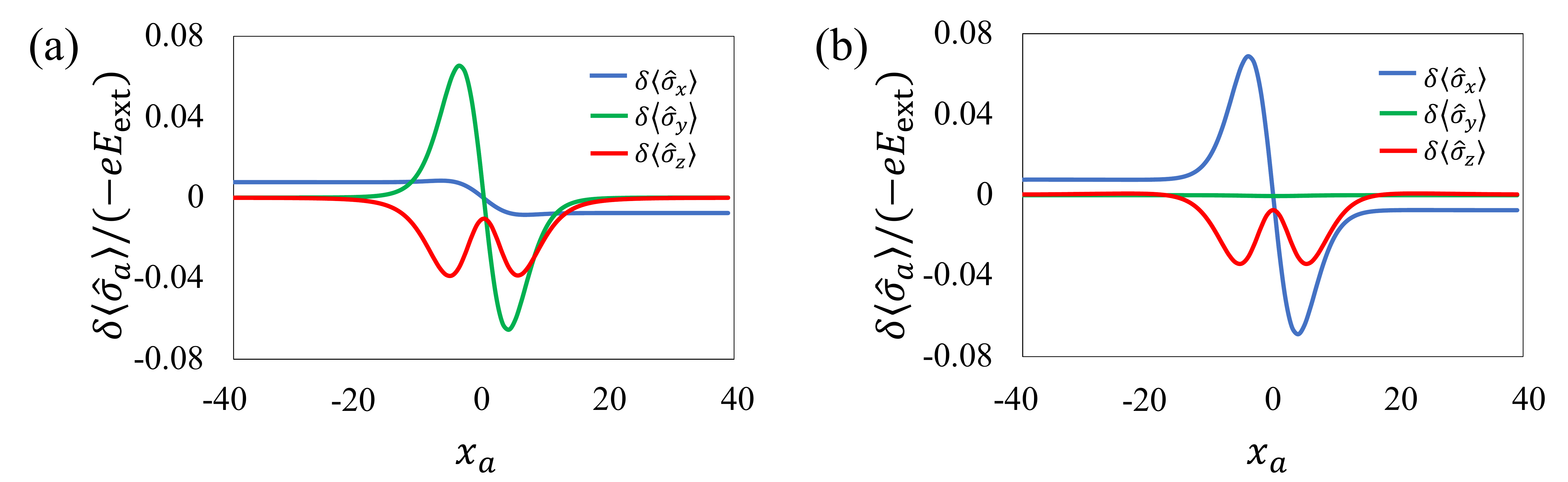}
\caption[]{(Color online) Numerical results of the spin accumulation \begin{math}\delta\langle\hat{\mbox{\boldmath $\sigma$}}(x)\rangle\end{math} under (a) Bloch-type domain wall and (b) N\'eel-type domain wall with the Rashba spin-orbit coupling (\begin{math}\lambda_{\mathrm{R}}/\lambda_{\mathrm{so}}=1\end{math}).  We choose the Hamiltonian parameters as \begin{math}\lambda_{\mathrm{so}} / t = 0.1\end{math}, \begin{math}\lambda_{\mathrm{R}}/t=0.1\end{math}, \begin{math}J_{\mathrm{exc}}S / t = 3.5\end{math}, and \begin{math}\xi_{a} = 30\end{math}. Fermi energy is \begin{math}E_{F}/t=-4.6\end{math}.} \label{rashba}
\end{figure*}
\indent In the above, we observe the non-equilibrium spin polarization \begin{math}\delta\langle\hat{\mbox{\boldmath $\sigma$}}(x)\rangle\end{math} which could drive the domain wall motion. We use the Thiele's approach, which describes that the magnetic texture moves at a constant velocity \begin{math}v\end{math} without deformation i.e. \cite{PhysRevLett.30.230} \begin{math}\hat{\bm{M}}=\hat{\bm{M}}(x-vt)\end{math}. In this case, the time derivative of \begin{math}\hat{\bm{M}}\end{math} is \begin{math}\frac{d\hat{\bm{M}}}{dt}=-v\partial_{x}\hat{\bm{M}}\end{math}. Therefore LLG equation becomes
\begin{equation} \label{eq:thiele01}
v\hat{\bm{M}}\times\partial_{x}\hat{\bm{M}}=\frac{J_{\mathrm{exc}}S}{\hbar\rho_{S}}\delta\langle\hat{\mbox{\boldmath $\sigma$}}(x)\rangle-\alpha v \partial_{x}\hat{\bm{M}}.
\end{equation}
By applying \begin{math}\int dx \partial_{x}\hat{\bm{M}}\end{math} to (\ref{eq:thiele01}), we obtain
\begin{equation} \label{eq:DWvelocity}
v=\frac{J_{\mathrm{exc}}S}{\hbar\rho_{S}}\frac{c_{2}}{c_{1}+c_{3}\alpha},
\end{equation}
where \begin{math}c_{1}=\int dx \partial_{x}\hat{\bm{M}}\cdot(\hat{\bm{M}}\times\partial_{x}\hat{\bm{M}})\end{math}, \begin{math}c_{2}=\int dx \partial_{x}\hat{\bm{M}}\cdot\delta\langle\hat{\mbox{\boldmath $\sigma$}}(x)\rangle \end{math} and \begin{math}c_{3}=\int dx \partial_{x}\hat{\bm{M}}\cdot\partial_{x}\hat{\bm{M}}\end{math}.
By the description of Eq. (\ref{eq:DWvelocity}), we estimate the domain wall velocity \begin{math}v\end{math}. Figure \ref{velo}(a) shows the domain wall velocity \begin{math}v\end{math} m/s as a function of the domain wall width divided by the lattice constant \begin{math}\xi_{a}\end{math}. According to the result, the domain wall velocity increases with the domain wall width. Note that our analysis might not be applicable with increasing the domain wall width because of an effect of a spin fluctuation\cite{PhysRevLett.30.230,TATARA2008213} and an edge reconstruction\cite{PhysRevB.49.8227}. Substituting concrete values into each parameter such as \begin{math}J_{\mathrm{exc}}S=0.2\end{math} eV, \begin{math}E_{\mathrm{ext}}=10^{5}\end{math} V/m, \begin{math}\xi=100\end{math} nm and physical quantities of \begin{math}\mbox{Co}_{3}\mbox{Sn}_{2}\mbox{S}_{2}\end{math}, the domain wall velocity is estimated as \begin{math}v\simeq520\end{math} m/s which is faster than that of conventional ferromagnetic nanowires \cite{PhysRevLett.98.037204,PhysRevLett.98.187202}.\\
\indent These results are different from those of the magnetic surface of the three-dimensional topological insulator which is also considered as the quantum anomalous Hall state. In the domain wall of the magnetic surface, the electron density distributions around the Bloch-type and the N\'eel-type domain walls are distinct from each other\cite{Wakatsuki_2015}. Thus, the domain wall velocity depends on the domain wall type. On the other hand, we can expect the same motion for both domain wall types in our model.\\
\indent In order to confirm whether the spin density \begin{math}\delta\langle\hat{\mbox{\boldmath $\sigma$}}(x)\rangle\end{math} is affected by the type of the spin-orbit coupling \cite{Hals_prb_spin} or not, we add the Rashba spin-orbit coupling (\begin{math}\lambda_{\mathrm{R}}/\lambda_{\mathrm{so}}=1\end{math}) to both Bloch-type [Fig. \ref{rashba}(c)] and N\'eel-type [Fig. \ref{rashba}(d)] domain walls. As a result, spatial distributions of \begin{math}\delta\langle\hat{\sigma}_{z}(x)\rangle\end{math} in both domain wall types are independent of domain wall types despite the additional spin-orbit coupling term whereas those of the perpendicular components \begin{math}\delta\langle\hat{\sigma}_{\perp}(x)\rangle\end{math} change according to the domain wall type. Thus, we assume the shape of \begin{math}\delta\langle\hat{\sigma}_{z}(x)\rangle\end{math} is caused by the quantum anomalous Hall state regardless of the type of spin-orbit coupling while the perpendicular component depends on it.\\
\indent Here we study the \begin{math}\delta\langle\sigma_{\perp}(x)\rangle\end{math} arising from the Rashba term. The Rashba spin-orbit coupling induces the additional topological contribution in the spin accumulation \begin{math}\delta\langle\hat{\mbox{\boldmath $\sigma$}}_{R}(x)\rangle\propto -\mbox{sgn}(M_{z})\bm{E}\end{math}. Focusing on the magnetic domain region where the magnetization is uniform, a finite spin accumulation corresponding to the effective magnetic field is generated. This is nothing but the spin-orbit torque which appears in single-domain ferromagnetic metals. \cite{PhysRevLett.92.126603,PhysRevB.79.094422} As distinct from the conventional a spin-orbit torque, however, the effective field is related to the Hall current perpendicular to the external electric field. The effective field is (anti-)parallel to the external electric field, changes their sign for the direction of local spins \begin{math}M_{z}\leftrightarrow -M_{z}\end{math}, and is realized in case of ferromagnetic insulating bulks. Garate and Franz \cite{PhysRevLett.104.146802} showed this effect at the interface between a topological insulator and a ferromagnetic thin layer. We obtain the same effect in the ferromagnetic Kagome lattice numerically.\\
\indent In summary, we investigate the electric field-induced spin torque on the domain wall on the ferromagnetic Kagome lattice in the quantum anomalous Hall regime. We find that the spin polarization of localized charges around domain walls is induced electrically and exerts non-adiabatic torques on the magnetic texture. We estimate the domain wall velocity within the Thiele's approach. The domain wall velocity becomes faster than that of conventional metals \cite{PhysRevLett.98.037204,PhysRevLett.98.187202}. Importantly, this phenomenon occurs even in the bulk insulator. Moreover, they indicate that the quantum anomalous Hall state has a potential to overcome a Joule heating problem. From these features related to high efficiency and low-energy consumption, we expect that quantum anomalous Hall states would be good candidates of a new type of spintronic devices.

\section*{Acknowledgement}
We would like to thank K. Kobayashi for helpful discussion. S.K. acknowledges support from GP-Spin at Tohoku University. D.K. was supported by the RIKEN Special Postdoctoral Researcher Program. This work was supported by JSPS KAKENHI Grants No. JP15H05854 and No. JP17K05485, and JST CREST Grant No. JPMJCR18T2.

\end{document}